\documentclass[a4paper,11pt]{article}
\usepackage[pdftex]{graphicx}
\usepackage{natbib}
\usepackage{geometry}
\title{Landing Dynamics of a Seagull Examined by Field Observation and Mathematical Modeling}
\author{Yuri Eisaki, Ikkyu Aihara, Isamu Hikosaka and Tohru Kawabe}

\begin{document}

\maketitle

\begin{abstract}
A seagull ({\it Larus crassirostris}) has a high ability to realize its safe, accurate and smooth landing. We examined how a seagull controls its angle of attack when landing on a specific target. First, we recorded the landing behavior of an actual seagull by multiple video cameras and quantified the flight trajectory and the angle of attack as time series data. Second, we introduced a mathematical model that describes how a seagull controls its speed by changing its angle of attack. Based on the numerical simulation combining the mathematical model and empirical data, we succeeded in qualitatively explaining the landing behavior of an actual seagull, which demonstrates that the control the angle of attack is important for landing behavior.
\end{abstract}

\section{Introduction}
In general, birds have a high ability of motion control. For example, hummingbirds hover at a fixed position when sucking the nectar of a flower \citep{soaring}; migratory birds fly a long distance when changing the place to live \citep{hovering}. To realize such a sophisticated motion control, birds recognize their surrounding environment through visual information and air flow, and adaptively change their position by flapping, gliding, turning and so on \citep{bird}.
Recently, the mechanism of the motion control is examined due to a high-definition field recording and mathematical modeling \citep{vicsek}. 
{\it Hayakawa} examined the adjustment mechanism of flock size in wild geese by comparing empirical data of video recordings with numerical simulation of a mean-field model \citep{hayakawa}; {\it Nagy et.al.} quantified the hierarchical flight dynamics of pigeons by analyzing empirical data of GPS tracking \citep{nagy}. 
Thus, empirical and theoretical studies contributed to further understandings of the motion control of flying birds.

In this study, we focus on the landing behavior of a seagull ({\it Larus crassirostris}).
A seagull inhabits a waterside, and is observed every season in Japan \citep{kamome}.
It is known that a seagull has long and narrow wings producing large lift \citep{bird} and then can land on a small target such as buoy by slightly controlling its angle of attack with less flapping even in a noisy actual environment.
Given that various species of birds and insects often flap just before landing \citep{hovering} \citep{bird}, it is expected that a seagull realizes a simple but sophisticated motion control required for landing.
Therefore, the study on the landing behavior of a seagull would contribute to the understandings of the flexible and highly accurate motion control of birds. 

The purpose of this study is to reveal the mechanism of the landing behavior, especially how a seagull controls its angle of attack just before landing.
First, we performed a field recording by using two video cameras and quantified the landing behavior of a seagull as time series data in a three-dimensional space (see Sec. \ref{chapter:data}).
Next, we theoretically analyzed the mechanism of the landing behavior based on numerical simulations of a mathematical model (equations of motion), and then compared its result with the empirical data (see Sec. \ref{chapter:model}).

\section{Materials and methods}
\subsection{Field observation}
\label{chapter:data}
In this section, we explain the method of a field recording and data analysis.
First, we define three angles that are related to the landing behavior of a seagull (see Sec. \ref{sec:aoa}).
Second, we explain the method of a field recording by using two video cameras (see Sec. \ref{sec:record}).
Third, we quantify the landing behavior by analyzing the video data (see Sec. \ref{sec:uma} and Sec. \ref{sec:DLT}).

\subsubsection{Definition of three angles related to landing behavior}
\label{sec:aoa}

\begin{figure}[h]
	\begin{center}
		\includegraphics[height=6cm]{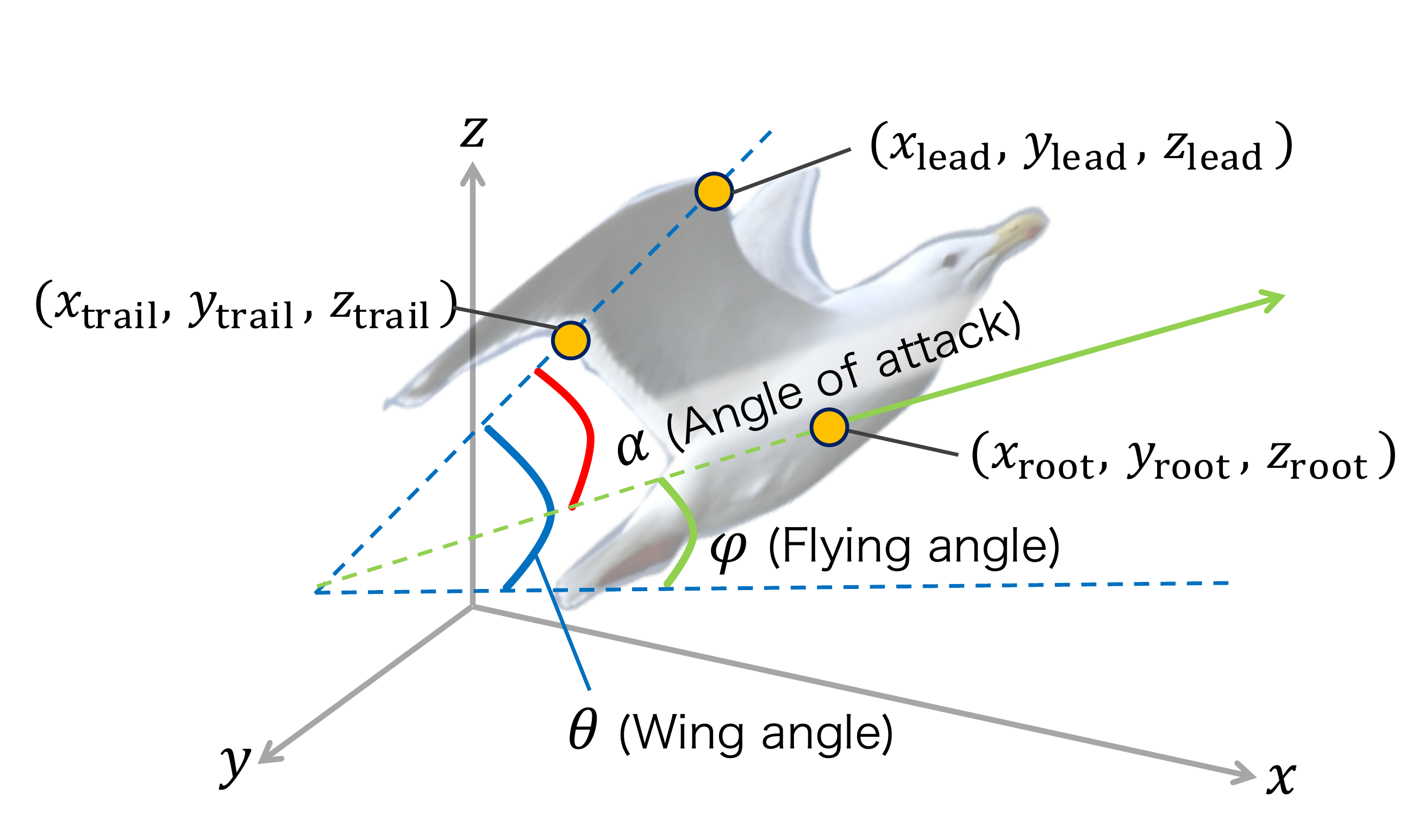}
		\caption{Definition of three angles (a wing angle ($\theta$), a flying angle ($\varphi$) and the angle of attack ($\alpha$)) in a three-dimension space.}
		\label{fig:mangle}
	\end{center}
\end{figure}

Here, we consider the situation in which a seagull flies in a three-dimensional space given by $x$-$y$-$z$ axis (see Fig. \ref{fig:mangle}).
Then, we focus on three angles (a wing angle, a flying angle and  the angle of attack) as the factors affecting the landing behavior of a seagull.
\vspace{-3mm}
\vspace{-3mm}

First, we define a wing angle $\theta$ as the angle between a chord line and the horizontal plane (i.e., the $x$-$y$ plane). 
A chord line represents the straight line connecting the leading edge of a wing with the trailing edge (see Fig. \ref{fig:mangle}). 
Then we describe a wing angle $\theta$ as follows:
\begin{eqnarray}
\theta(i)&=&\arctan\frac{z_{\rm lead}(i)-z_{\rm trail}(i)}{\sqrt{({x_{\rm lead}(i)-x_{\rm trail}(i)})^2 + ({y_{\rm lead}(i)-y_{\rm trail}(i))^2}}}.
\label{theta}
\end{eqnarray}
Here, ($x_{\rm lead}(i)$, $y_{\rm lead}(i)$, $z_{\rm lead}(i)$) is the position of the leading edge of the wing at the $i$th frame of a video data in the three-dimension space, and ($x_{\rm trail}(i)$, $y_{\rm trail}(i)$, $z_{\rm trail}(i)$) is the position of the trailing edge of the wing.
We calculate Eq. (\ref{theta}) by using the function of {\it atan2} that is provided by MATLAB, and then obtain the wing angle $\theta$ as the variable ranging from $-\pi$ to $\pi$ [rad] (mod $2\pi$).

Next, we define a flying angle $\varphi$ as the angle between the velocity in the vertical axis (the $z$ axis) and the velocity in the horizontal plane (the $x$-$y$ plane). 
Then, we describe the flying angle $\varphi$ as follows: 
\begin{eqnarray}
\varphi(i)=\arctan\frac{z_{\rm root}(i)-z_{\rm root}(i-1)}{\sqrt{({x_{\rm root}(i)-x_{\rm root}(i-1))^2 + (y_{\rm root}(i)-y_{\rm root}(i-1))^2}}}.
\label{eq:idoeq}
\end{eqnarray}             
Here, ($x_{\rm root}(i)$, $y_{\rm root}(i)$, $z_{\rm root}(i)$) is the position of the root of the wing at the $i$th frame in the three-dimensional space, and ($x_{\rm root}(i-1)$, $y_{\rm root}(i-1)$, $z_{\rm root}(i-1)$) is the position at the $(i-1)$th frame. 
We calculate Eq. (\ref{eq:idoeq}) by using the function of {\it atan2} as the same with the case of the wing angle, and then obtain the flying angle $\varphi$ as the variable ranging from $-\pi$ to $\pi$ [rad] (mod $2\pi$).

Finally, we calculate the angle of attack $\alpha$ as the angle between the wing angle $\theta$ and the flying angle $\varphi$ as follows:
\begin{eqnarray}
\alpha(i)&=&\theta(i)-\varphi(i).\label{eq:kakudo1}
\end{eqnarray}
Here, we define the angle of attack $\alpha$ as the variable ranging from $-\pi$ to $\pi$ [rad] (mod $2\pi$), which is consistent with the definitions of the wing angle and the flying angle.
It is well known that the angle of attack $\alpha$ dominantly affects the forces (lift and drag) associated with a flying bird (see Sec. \ref{sec:modelsiki}).
Therefore, we focus on the dynamics of the angle of attack in this study for both of the field observation and mathematical modeling.

\subsubsection{ Field recording }
\label{sec:record}

\begin{figure}[htb]
	\begin{center}
		\includegraphics[height=5.5cm]{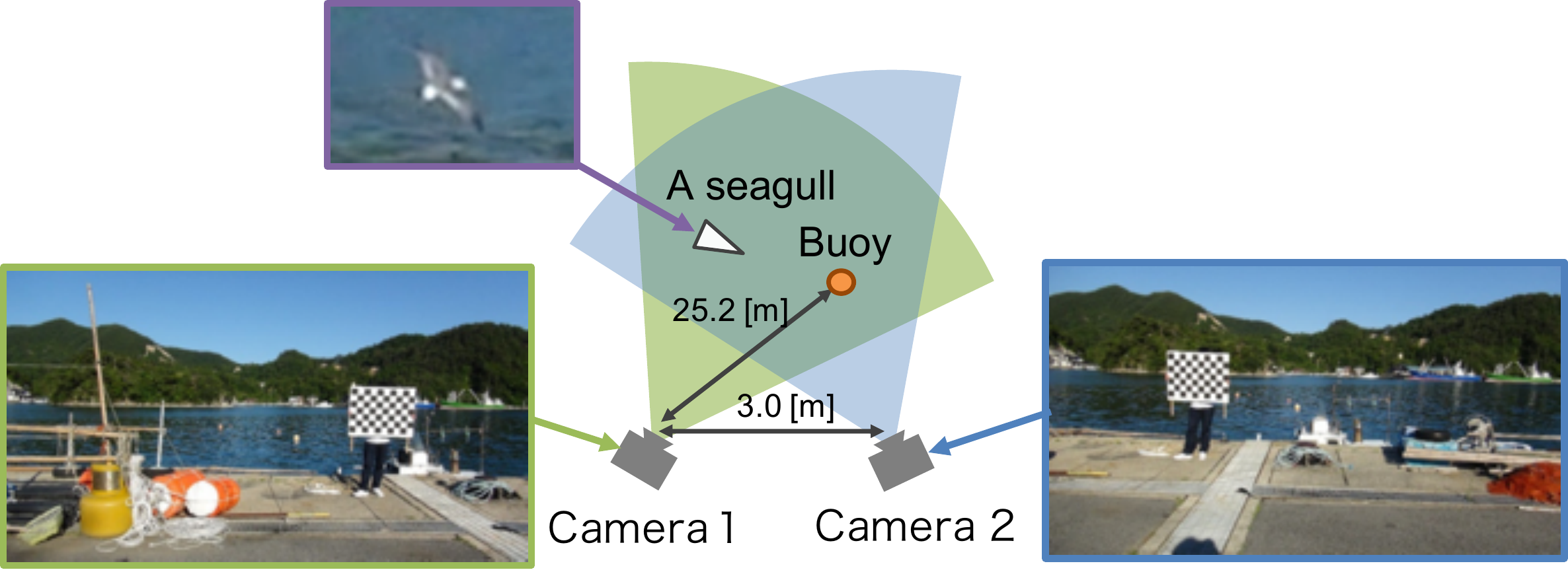}
		\caption{Setup of our field recording.}
		\label{fig:jikken}
	\end{center}
\end{figure}

In an actual field, we recorded the landing behavior of a seagull.
As shown in Fig. \ref{fig:jikken}, we set two video cameras (SONY HDR-XR550V) at different angles and then recorded the seagull that landed on a small target point (buoy) floating on a sea.
The frame rate of the cameras was set at 59.940 [frame/second], and the pixel size of each video frame was set at 1920$\times$1080 [pixels].
The field recording was carried out at 16:35 on 4th June, 2017, at Oki island, Shimane prefecture, Japan.
The weather was fine; the temperature and humidity were 20.1 [$^\circ$C] and 57.6 [\%], respectively. 

Next, we recorded a set of pictures for determining calibration parameters of video cameras.
For this recording, we made a checkerboard of 80.0$\times$110.0 [cm] on which the 7$\times$10 grids of 9.05$\times$9.05 [cm] were printed (see Fig. \ref{fig:jikken}), and then took 20 pictures of the checkerboard that were placed at slightly different positions. 
Note that the multiple video data and calibration parameters are necessary for quantifying the landing behavior of the flying seagull (see Sec. \ref{sec:uma}) and reconstructing its dynamics in the three dimension space (see Sec. \ref{sec:DLT}).

\vspace{-3mm}

\subsubsection{Tracking of a flying seagull}
\label{sec:uma}
\begin{figure}[h]	
	\begin{minipage}[b]{0.5\hsize}
		{\huge A.} 
		\begin{center}
			\includegraphics[height=4cm]{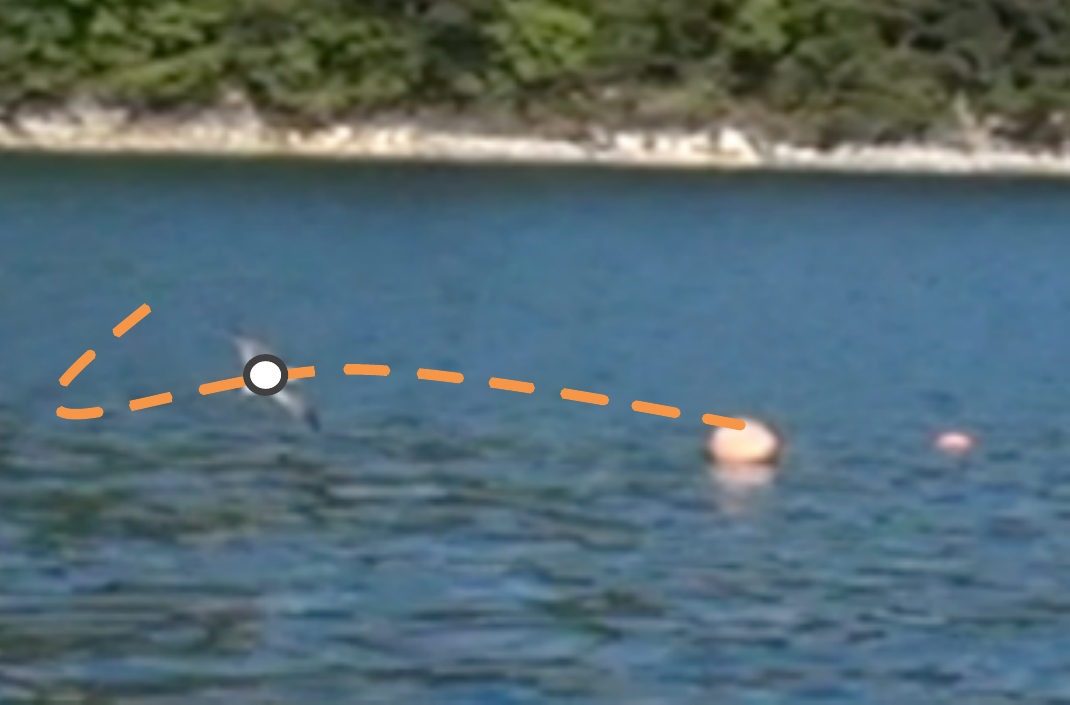}
		\end{center}
	\end{minipage}
	\begin{minipage}[b]{0.5\hsize}
		{\huge B.} 
		\begin{center}
			\includegraphics[height=4cm]{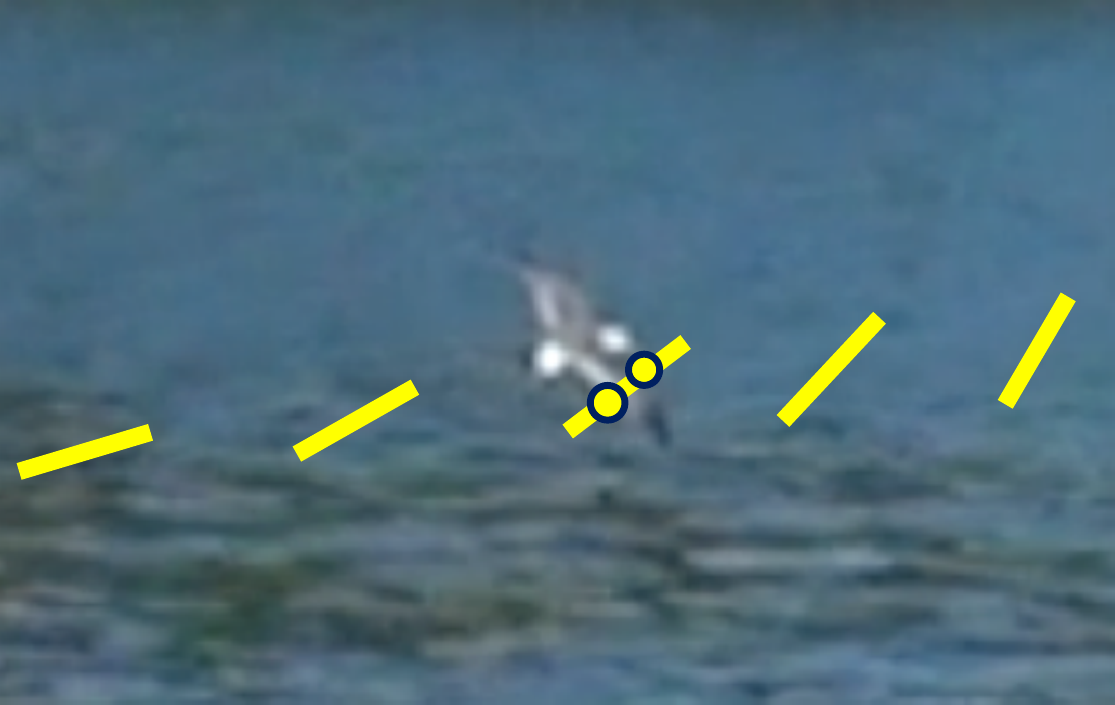}
		\end{center}
	\end{minipage}
	\caption{ Quantification of the landing behavior of a seagull. We tracked (A) the flight trajectory and (B) the wing angle from multiple video data by using a software UMATracker.}
	\label{fig:uma}
	
\end{figure}

To estimate the flight trajectory and the wing angle, we analyzed the video data by using the software UMATracker \citep{umatracker}.
This software allows us to automatically track the position of a moving object from a video data, and is widely applied to the study on the behavior of various animals \citep{umatracker}.
In this study, we tracked the root of a wing of a flying seagull (see Fig. \ref{fig:uma}A) and the chord line (see Fig. \ref{fig:uma}B) frame by frame by using the software, which corresponds to the quantification of the flight trajectory and the wing angle.
This analysis is carried out for video data of two cameras, respectively.

\subsubsection{Reconstruction of the landing behavior in a three dimensional space}
\label{sec:DLT}
We reconstructed the landing behavior of a flying seagull in the three dimensional space by the Direct Linear Transformation method (the DLT method) \citep{DLT}.
The DLT method is a well-known framework for converting a position recorded by multiple cameras into the coordinates in a three dimensional space, but
we need to determine the parameters of the conversion prior to applying the DLT method \citep{DLT}.
In this study, we estimated the parameters of the conversion by using the pictures of the checkerboard that were obtained from our field recording (see Sec. \ref{sec:record}).
Then, we reconstructed the flight trajectory and the wing angle in the three-dimensional space by analyzing the video data of two cameras according to the DLT method on the assumption of the estimated parameters of the conversion.

In addition, we smoothed the time series data of the flight trajectory and the wing angle by spline interpolation so as to reduce the error of estimation.

\subsection{Mathematical modeling of the landing behavior}
\label{chapter:model}

\begin{figure}[h]
	\begin{center}
		\includegraphics[height=5.5cm]{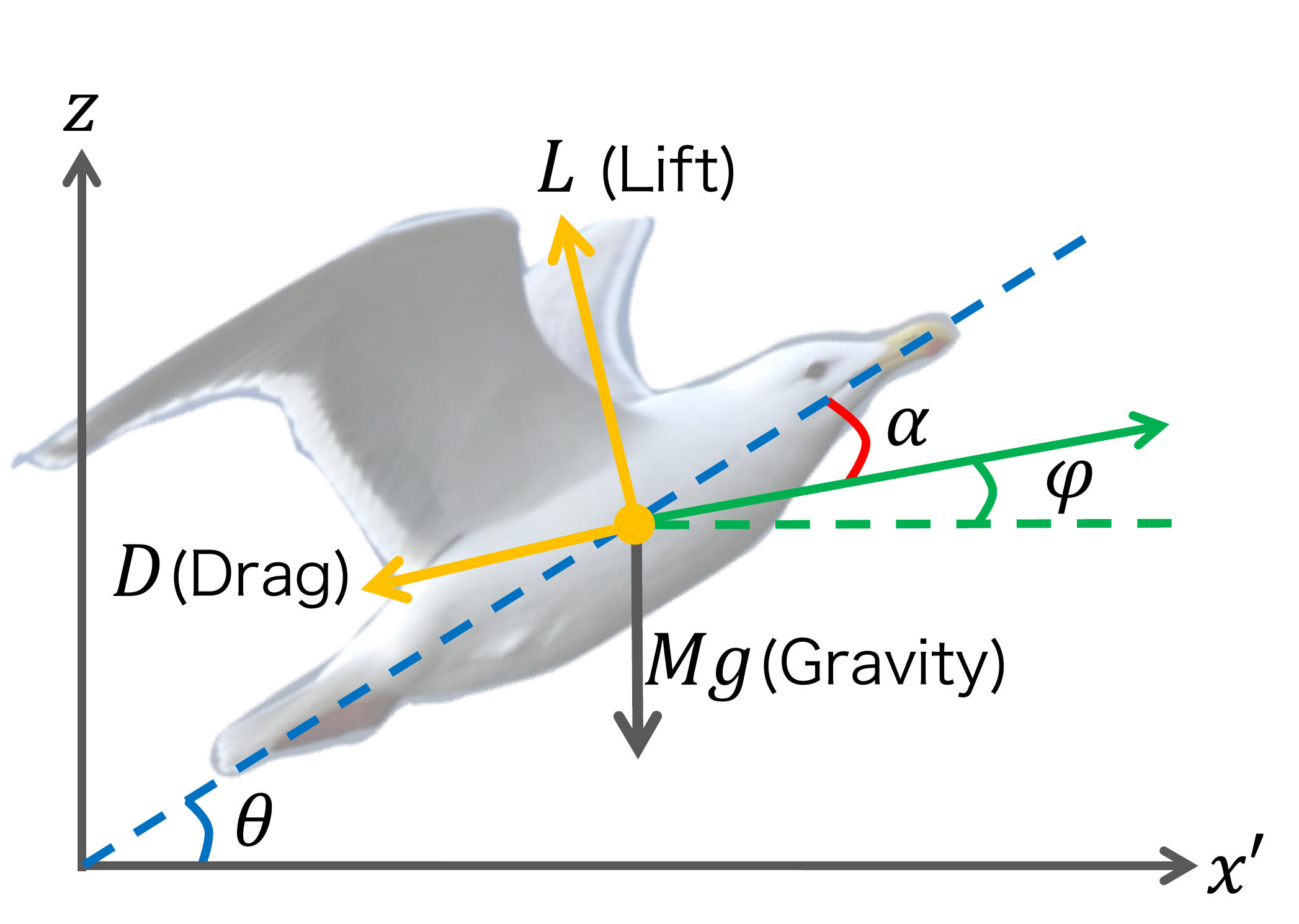}
		\caption{Three kinds of forces associated with the flight of a seagull.}
		\label{fig:model}
	\end{center}
\end{figure}

In our field recording, we confirmed that the seagull did not wobble in the horizontal plane just before landing.
Therefore, we define  $x'$-axis as the axis along the flight direction in the horizontal plane, and consider the situation in which the seagull flies in the same vertical plane described by the $x'$-axis and $z$-axis (see Fig. \ref{fig:model}).
Then, the flight dynamics of the seagull is modeled by equations of motions as follows \citep{aoa}:
\begin{eqnarray}
M\frac{dv_{x'}}{ dt}&=&-D(\alpha)\cos(\varphi)-L(\alpha)\sin(\varphi),\label{eq:Max}\\
M\frac{dv_{z}}{ dt}&=&L(\alpha)\cos(\varphi)-D(\alpha)\sin(\varphi)-Mg.\label{eq:Maz}
\end{eqnarray}
Here, $v_{x'}$ and $v_z$ describe the velocity in the $x'$-axis and $z$-axis, respectively. 
The parameter $M$ is the mass of the seagull, and the parameter $g$ is the gravitational acceleration.
Then, the functions $L(\alpha)$ and $D(\alpha)$ represent the forces of lift and drag that act on the seagull vertically and horizontally to the flight direction $\varphi$ \cite{LD,aoa}.
We describe these forces as follows :
\begin{eqnarray}
L(\alpha)&=&\frac{1}{2}\rho S({v_{x'}}^2+{v_z}^2 ) C_l (\alpha),\label{eq:cl}\\
D(\alpha)&=&\frac{1}{2}\rho S({v_{x'}}^2+{v_z}^2 ) C_d (\alpha).\label{eq:cd}
\end{eqnarray}
The parameter $\rho$ is the mass density of air, and the parameter $S$ is the area of two wings.
In addition, $C_l(\alpha)$ and $C_d(\alpha)$ give the coefficients of lift and drag as the functions of the angle of attack $\alpha$.
\vspace{-3mm}

\begin{figure}[h]
	\begin{center}
		\includegraphics[height=4.5cm]{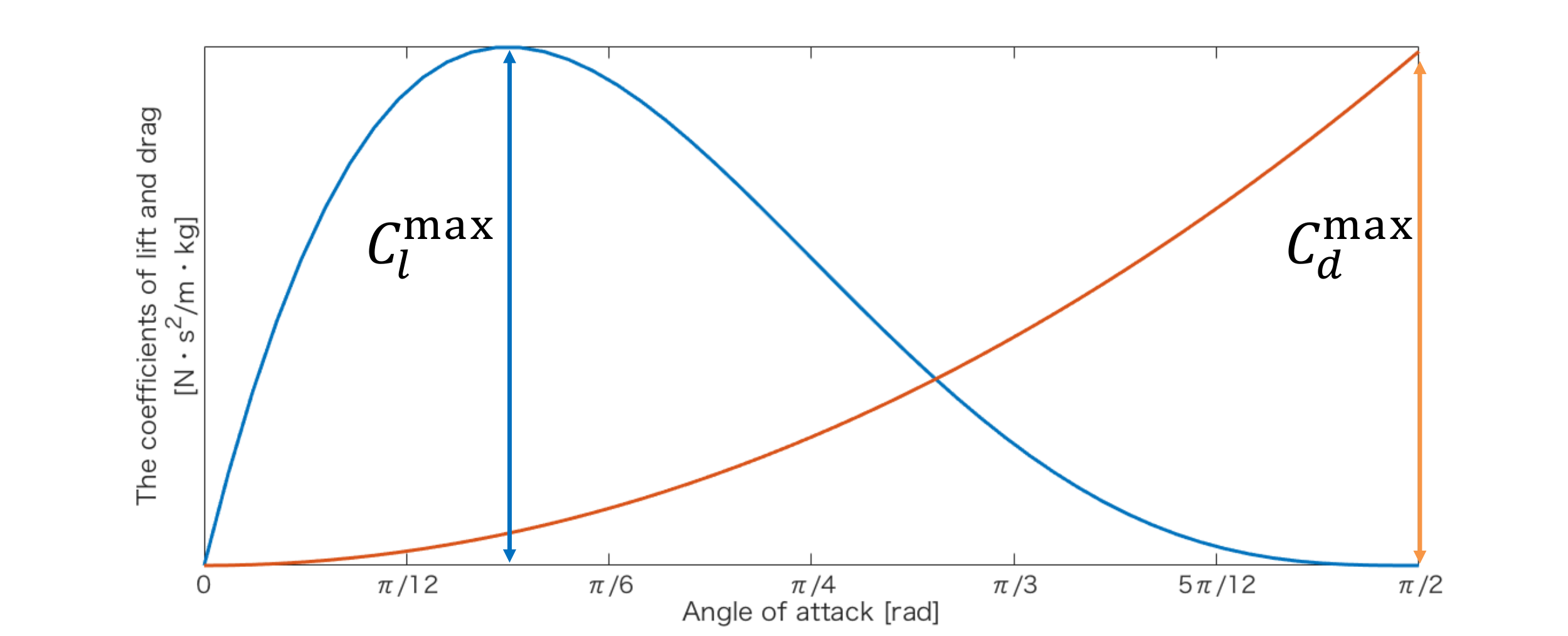}
		\caption{The formulation of the coefficients $C_l(\alpha)$ and $C_d(\alpha)$ that are related to lift and drag acting on a flying bird. The blue line represents the coefficient of lift  $C_l(\alpha)$, and the orange line represents the coefficient of drag  $C_d(\alpha)$.}
		\label{fig:LD}
	\end{center}
\end{figure}

It is known that the coefficients $C_l(\alpha)$ and $C_d(\alpha)$ strongly depend on the aerodynamic characteristics around the wings of a flying bird \citep{LD}.
The previous study \citep{LD} reported that the coefficient $C_l(\alpha)$ increases until the angle of attack reaches a threshold (almost $\pi/8$ in the case of a flying bird); $C_l(\alpha)$ start decreasing when the angle of attack exceeds the threshold because of the loss of airflow.
On the other hand, the coefficient $C_d(\alpha)$ increases monotonously as the angle of attack increases.
Based on these characteristics, we formulate $C_l(\alpha)$ and $C_d(\alpha)$ as shown in Fig. \ref{fig:LD} by using two parameters  $C_l^{\rm max}$ and $C_d^{\rm max}$.

Note that these framework is based on the previous study \citep{LD}, and we utilize these framework for the analysis on our empirical data of an actual seagull (Sec. \ref{sec:simkekka}).

\section{Results}
\subsection{Empirical Data}
\label{sec:datakekka}
From our field recording and data analysis on a flying seagull (Sec. \ref{chapter:data}), we obtained the flight trajectory and the angle of attack in the three-dimensional space.

\begin{figure}[h]
	\begin{center}
		\includegraphics[height=6cm]{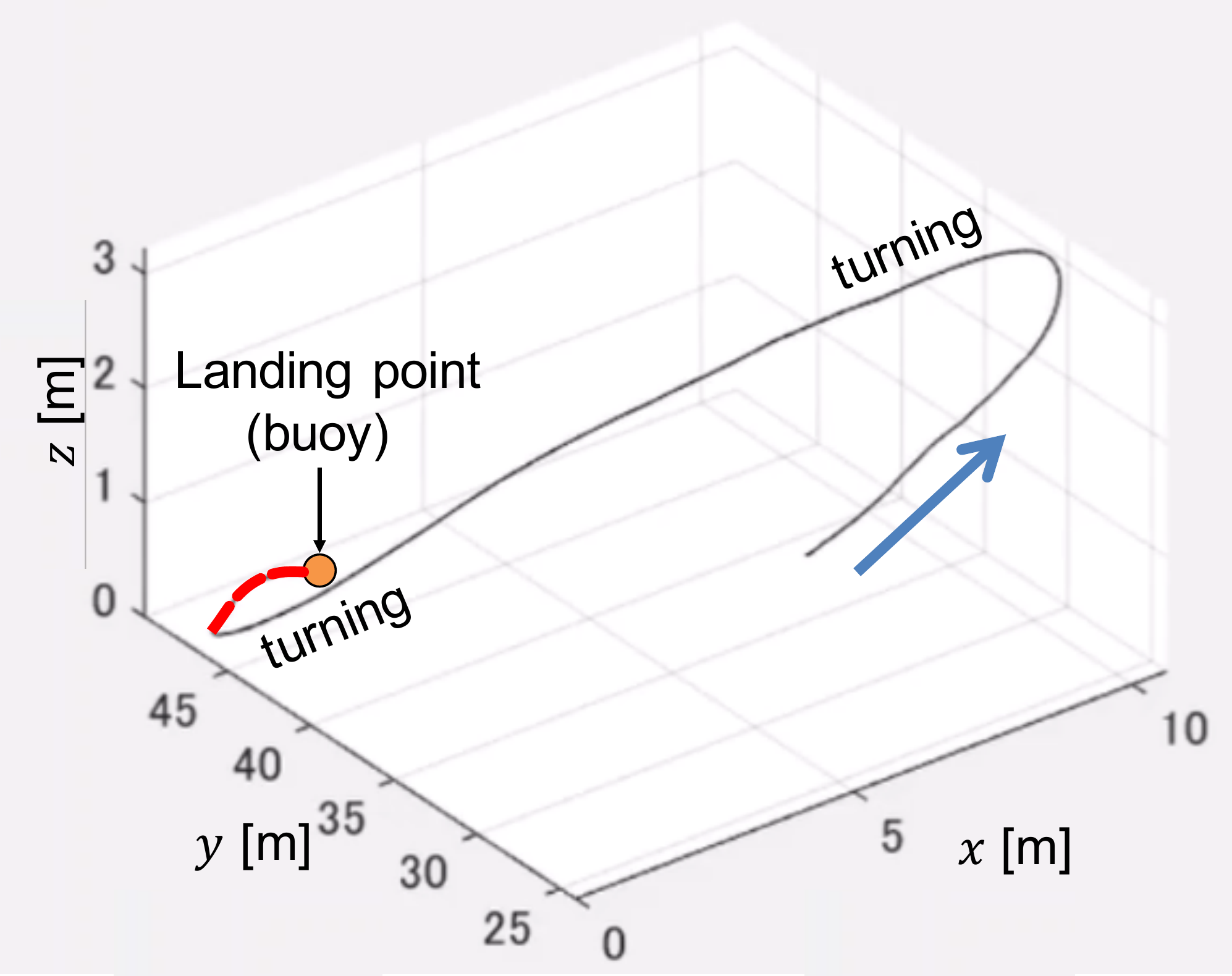}
		\caption{The flight trajectory of a seagull obtained from our field recording. In this data, a seagull succeeded in landing on the buoy.}
		\label{fig:tra}
	\end{center}
\end{figure}

Figure \ref{fig:tra} shows the flight trajectory of the seagull in eight seconds before landing. This data demonstrated that the seagull turned twice and successfully landed on the buoy.
During this flight, the seagull moved 25.2 [m] and fell 3.2 [m] in total.
We confirmed that the seagull did not turn and flap for 1.50 seconds before landing.
Because we do not consider the effect of turning and flapping in our mathematical model, we focus on this empirical data of 1.50 seconds before landing (see red line in Fig. \ref{fig:tra}).

\begin{figure}[h]
	\begin{center}
		\includegraphics[height=5cm]{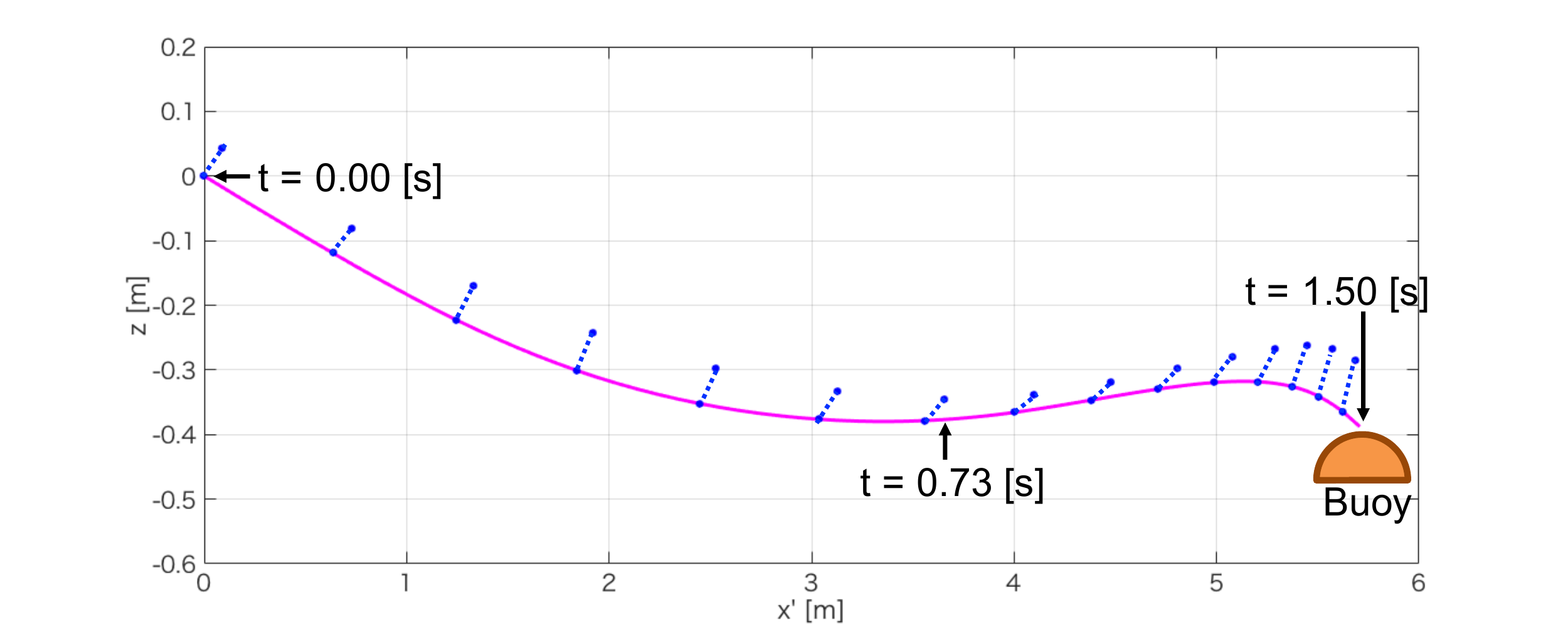}
		\caption{Flight trajectory and the wing angle in 1.50 seconds before landing. This data is obtained from our field recording.}
		\label{fig:datatra}
	\end{center}
\end{figure}

Figure \ref{fig:datatra} shows the flight trajectory and the wing angle of the seagull in 1.50 seconds before landing.
In this figure, the pink line represents the flight trajectory of a seagull, and the blue line represents the wing angle of the seagull in every 0.10 seconds.
The seagull flew downwards until 0.73 seconds, and then increased the wing angle.
Consequently, the seagull slightly flew upward, and landed on the buoy.

\begin{figure}[h]
	\begin{center}
		\includegraphics[height=5cm]{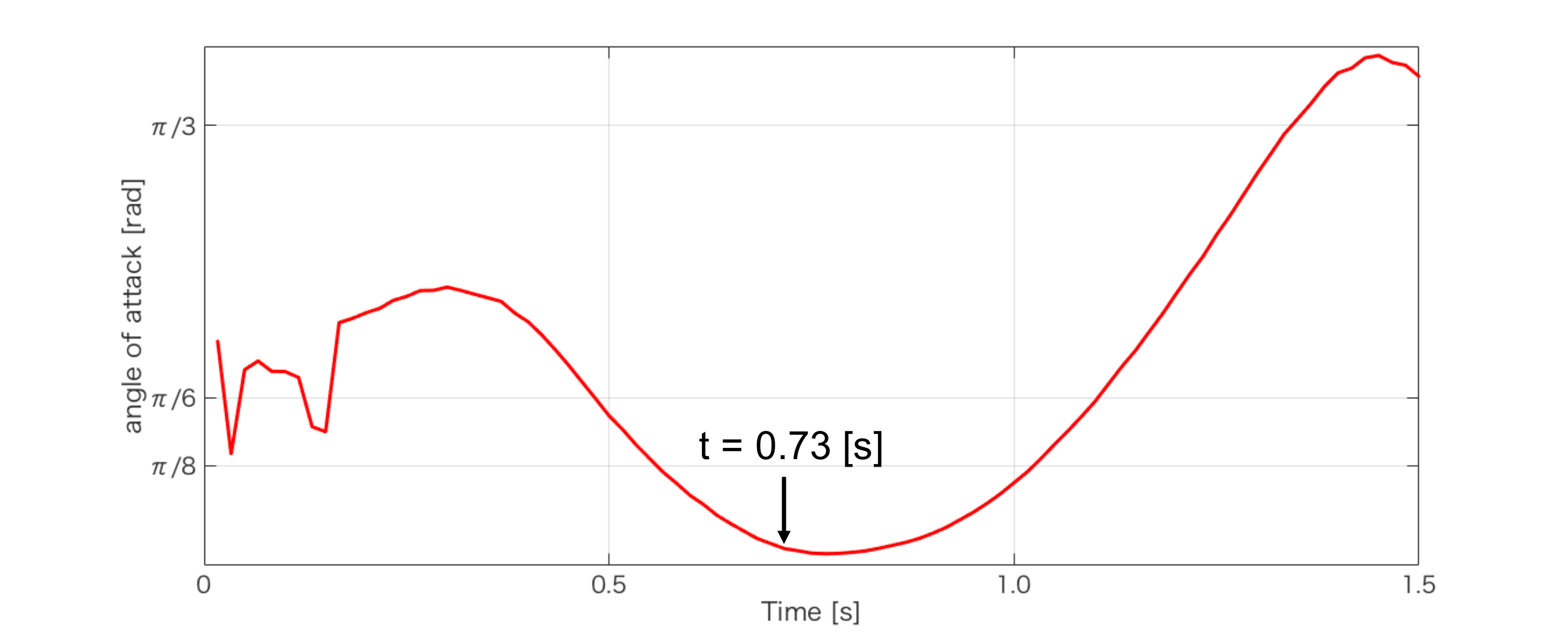}
		\caption{Time series data of the angle of attack $\alpha$ in 1.50 seconds before landing.	This data is obtained from our field recording.}
		\label{fig:angle}
	\end{center}
\end{figure}

Figure \ref{fig:angle} shows the time series data of the angle of attack $\alpha$ in 1.50 seconds before landing. 
The seagull controlled its angle of attack around $\pi/8$ until 0.73 seconds.
Given the shapes of lift and drag shown in Fig. \ref{fig:LD}, this property allows the seagull to keep large lift with small drag that is necessary for avoiding stall and keep flying.
Then, the seagull drastically increased its angle of attack.
This property allows the seagull to produce large drag that is essential for landing (see Fig. \ref{fig:LD}).

\subsection{Numerical simulation based on empirical data and mathematical model}
\label{sec:simkekka}
We perform numerical simulation to examine the mechanisms of the landing behavior based on the empirical data and the mathematical model.
By substituting the time series data of the angle of attack $\alpha$ (see Fig. \ref{fig:angle}) into the mathematical model (Sec. \ref{chapter:model}) and performing numerical simulation, we examined how the angle of attack $\alpha$ determines the flight dynamics of the seagull.
First, we fix the parameters of the mathematical model from field observation and previous studies (Sec. \ref{para}).
Second, we explain the result of numerical simulation (Sec. \ref{behavior}).
Third, we discuss the efficacy of the flight dynamics of an actual seagull by comparing the result of the numerical simulation with the empirical data (Sec. \ref{compare}).

\subsubsection{Parameter values}
\label{para}
To perform numerical simulation, we fix the parameters of the mathematical model (Eqs. (\ref{eq:Max})-(\ref{eq:cd})) as shown in Table \ref{tb:parameter}.
The area of wings $S$ and the mass of a seagull $M$ (see Eqs. (\ref{eq:cl}) and (\ref{eq:cd})) are fixed from the measurement of actual seagulls reported in a previous study \citep{birddata}.
Note that we cannot find the parameter values of {\it Larus crassirostris} (the target species of this study) and therefore utilize the parameter values of the similar species {\it Larus canus}.
The mass density of air $\rho$ and the gravitational acceleration $g$ are fixed at typical values.
The initial conditions of $v_{x'}$ and $v_z$ (i.e., $v_{{x'}_0}$ and $v_{z_0}$ in Table \ref{tb:parameter}) are calculated from our empirical data shown in Fig. \ref{fig:datatra}.

\begin{table}[h]
	\centering
	\caption{Parameter values used for numerical simulation.}
	\begin{tabular}{|l|l|l|}\hline
		$S$ & Area of wings & 0.115 [m$^2$]\\\hline
		$M$ & Mass of a seagull & 0.370 [kg]\\\hline
		$\rho$ & Mass density of air & 1.205 [g/m$^3$]\\\hline
		$g$ & Gravitational acceleration & 9.806 [m/s$^2$]\\\hline
		$v_{{x'}_0}$ & Initial velocity in $x'$ axis & 5.3 [m/s]\\\hline
		$v_{z_0}$ & Initial velocity in $z$ axis & -0.2 [m/s]\\\hline
		
	\end{tabular}
	\label{tb:parameter}
\end{table}
It is well known that the coefficients of lift and drag (e.g., $C_l^{\rm max}$ and $C_d^{\rm max}$) depend on the flexibility, symmetry and density of wings that can vary among the species of birds \citep{ClCd}.
However, to our knowledge, the measurement of $C_l^{\rm max}$ and $C_d^{\rm max}$ has not been conducted in the case of a seagull.
Therefore, we perform numerical simulation by varying these parameters in the range of $0.0\sim10.0$ with the interval of 0.25, respectively. Note that this range includes  $C_l^{\rm max}$ and $C_d^{\rm max}$ measured in other species of birds \cite{ClCd}.  

On the assumption of the above parameter values, we numerically calculate Eqs. (\ref{eq:Max}) and (\ref{eq:Maz}) due to Runge-Kutta method for 1.50 seconds with the time step of 0.00015 seconds.

\subsubsection{Numerical simulation}
\label{behavior}
\begin{figure}[h]
	\begin{center}
		\begin{tabular}{cc}
			\begin{minipage}{0.5\hsize}
				{\huge A.} 
				\begin{center}
					\includegraphics[height=3cm]{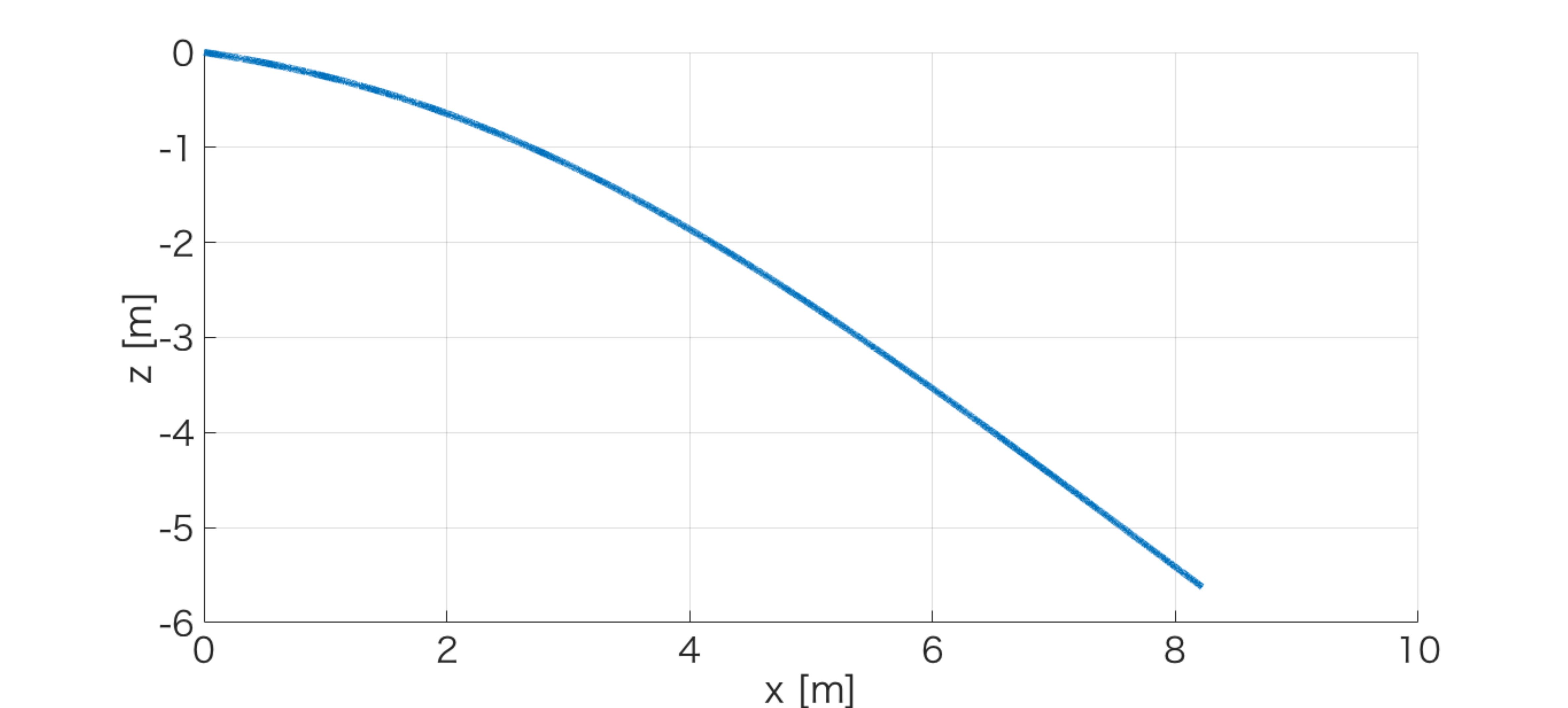}
				\end{center}
			\end{minipage}
			\begin{minipage}{0.5\hsize}
				{\huge B.} 
				\begin{center}
					\includegraphics[height=3cm]{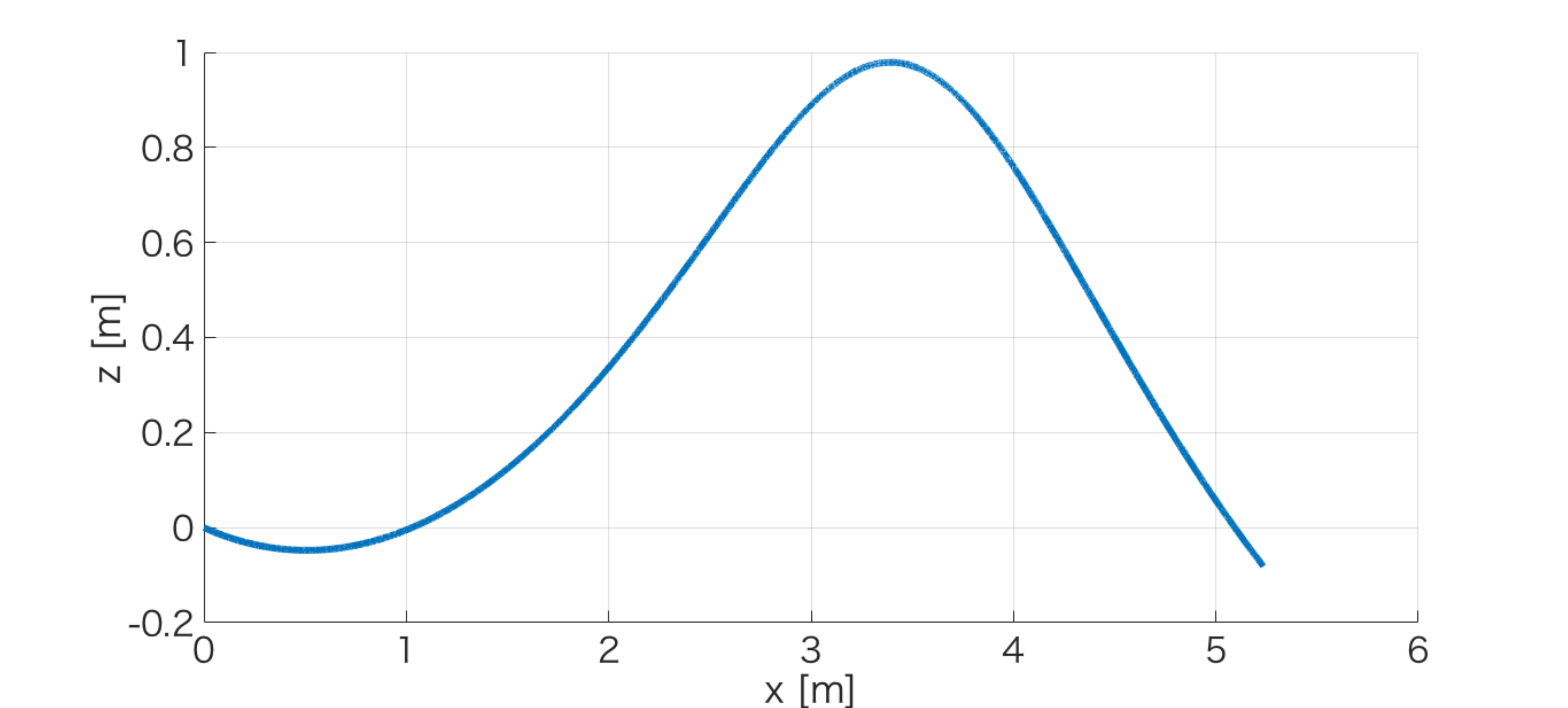}
				\end{center}
			\end{minipage}\\
			\begin{minipage}{0.5\hsize}
				{\huge C.} 
				\begin{center}
					\includegraphics[height=3cm]{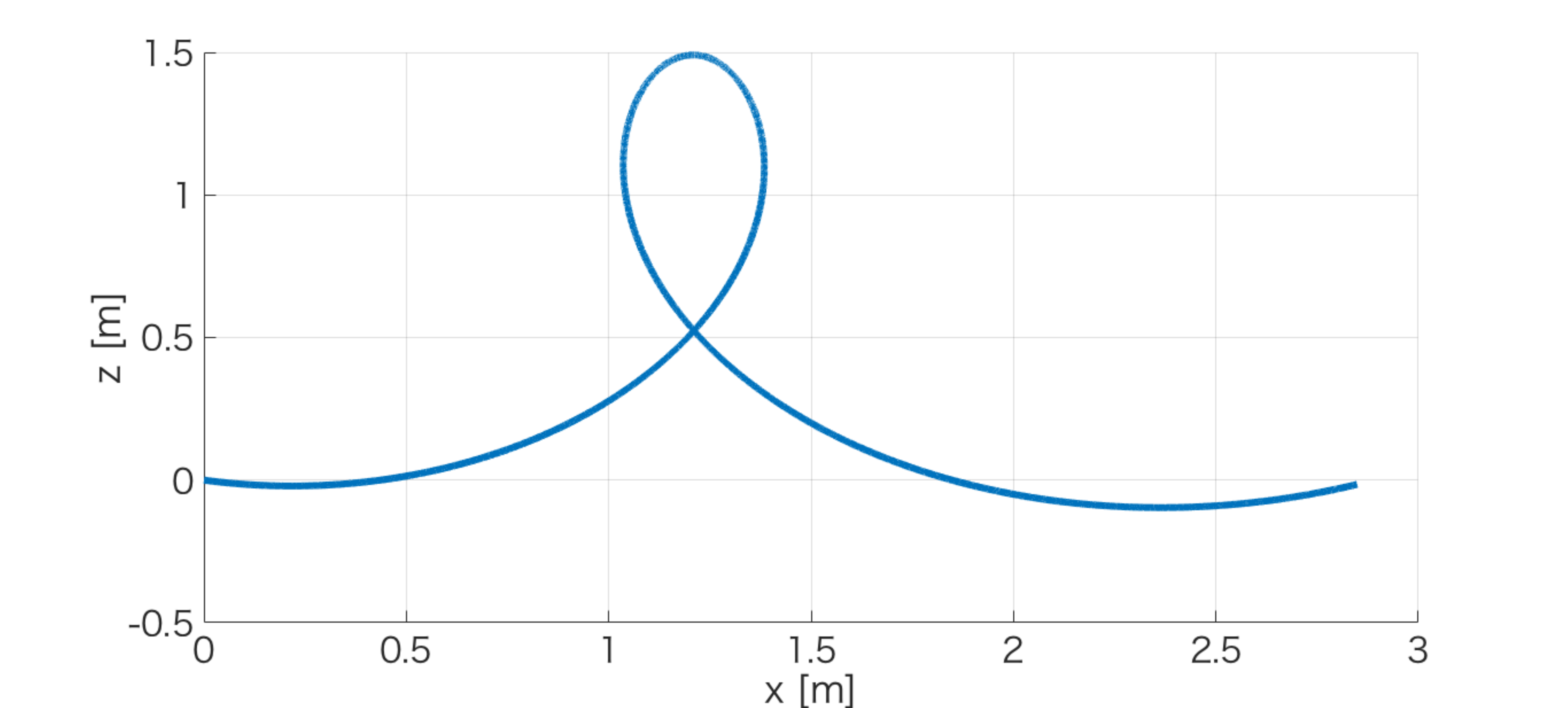}
				\end{center}
			\end{minipage}
			\begin{minipage}{0.5\hsize}
				{\huge D.} 
				\begin{center}
					\includegraphics[height=3cm]{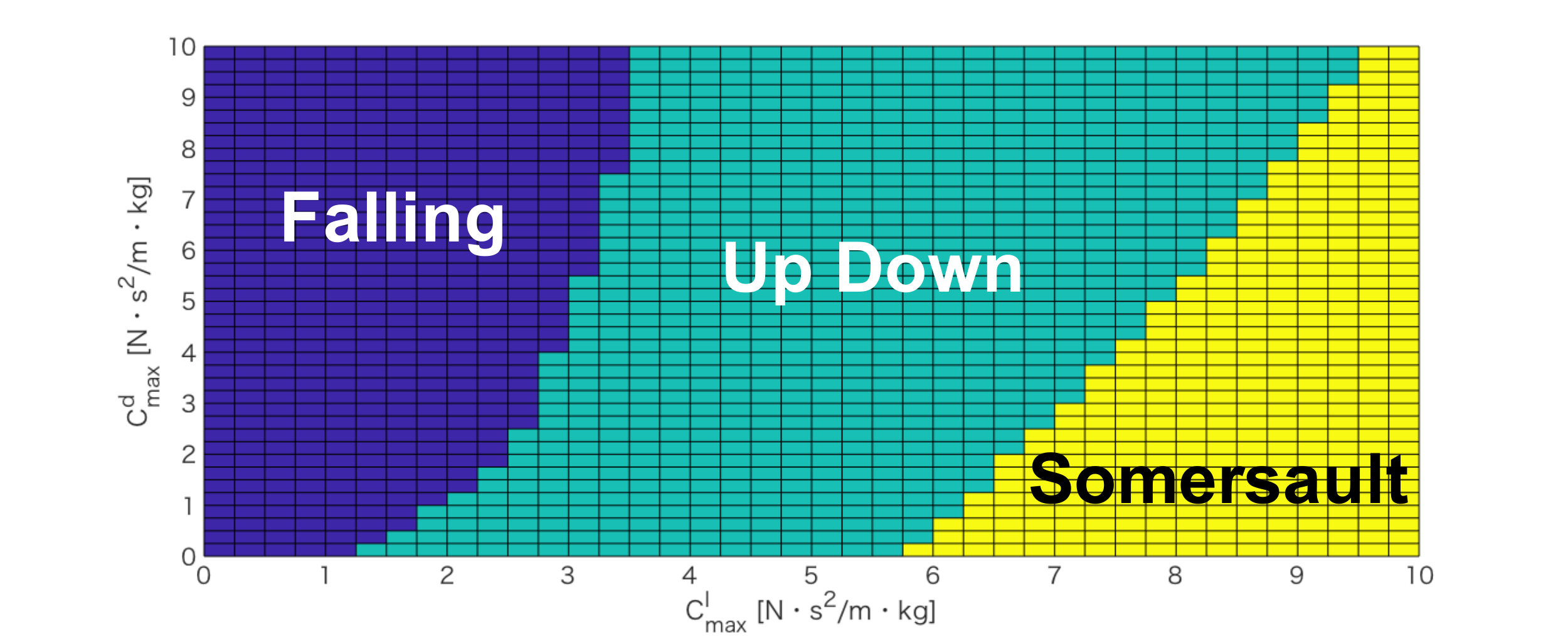}
				\end{center}
			\end{minipage}
		\end{tabular}
		\caption{Three types of the flight patterns and its bifurcation diagram. (A) An example of the flight pattern "falling" with parameter values ($C_l^{\rm max}$, $C_d^{\rm max}$) = (1.0, 5.0). (B) An example of the flight pattern "up and down" with parameter values ($C_l^{\rm max}$, $C_d^{\rm max}$) = (5.0, 5.0). (C) An example of the flight pattern "somersault" with parameter values ($C_l^{\rm max}$, $C_d^{\rm max}$) = (9.0, 5.0). (D) Bifurcation diagram among the three types of the flight patterns. Note that the type (B) is similar to the trajectory of our empirical data.}
		\label{fig:modelkekka}
	\end{center}
\end{figure}

By varying the values of $C_l^{\rm max}$ and $C_d^{\rm max}$, we perform numerical simulation of the mathematical model (Eqs. (\ref{eq:Max}) and (\ref{eq:Maz})) and obtained various trajectories.
We categorize the trajectories into three types of flight patterns, "falling", "up and down", and "somersault".
The first pattern "falling" means that a seagull monotonously moves downward as shown in Fig. \ref{fig:modelkekka}A; we detect this pattern when the condition of $v_z<0$ always holds for 1.50 seconds. 
The second pattern "up and down" means that a seagull repeats rising up and falling down as shown in Fig. \ref{fig:modelkekka}B; we detect this pattern when the condition of $v_{x'}\geq0$ holds and $v_z$ takes a positive value at least once.
The third pattern "somersault" means that a seagull rotate in the vertical plane as shown in Fig. \ref{fig:modelkekka}C; we detect this pattern when $v_{x'}$ takes a negative value at least once. 

Figure \ref{fig:modelkekka}D shows the bifurcation diagram among the three types of the flight patterns.
In this figure, the vertical axis represents $C_l^{\rm max}$, and the horizontal axis represents $C_d^{\rm max}$.
Each set of ($C_l^{\rm max}$, $C_d^{\rm max}$) is drawn with different colors describing the three types of the flight patterns; a blue region represents the flight pattern "falling", a green region represents the flight pattern "up and down", and a yellow region represents the flight pattern "somersault".
The point is that the bifurcation diagram includes a wide region of the flight pattern "up and down" and the pattern is consistent with the trajectory of our empirical data (see Fig. \ref{fig:datatra}).
This indicates that the control of the angle of attack $\alpha$ by an actual seagull is effective for realizing the flight pattern in which a seagull repeats rising up and falling down just before landing.

\subsection{Comparison between numerical simulation and empirical data}
\label{compare}

In this section, we search a suitable set of $(C_{l}^{\rm max}$, $C_{d}^{\rm max})$ that reproduces the flight trajectory of our empirical data (see Fig. \ref{fig:datatra}) as precisely as possible.
To quantify the difference between the numerical simulation and empirical data, we define the following evaluation function:
\begin{eqnarray}
J(x'_{\rm data}, z_{\rm data}, x'_{\rm model}, z_{\rm model})&=&\sum_{i=1}^{I} \sqrt{(z_{\rm data}(i)-z_{\rm model}(k))^2}.\label{eq:grid1}
\end{eqnarray}
Here, ($x'_{\rm data}(i)$, $z_{\rm data}(i)$) is the position of the seagull at the $i$th frame of the empirical data (see Fig. \ref{fig:datatra}).
The parameter $I$ is the total number of frames of the empirical data.
Then, ($x'_{\rm model}(k)$, $z_{\rm model}(k)$) is the position of the seagull at the $k$th step of the numerical simulation.
Note that the index $k$ is chosen as the value minimizing the difference  between $x'_{\rm model}(m)$ and $x'_{\rm data}(i)$ ($m=1,2,3,..,k,...$).
%
%

\begin{figure}[h]
	\begin{center}
		\includegraphics[height=5.5cm]{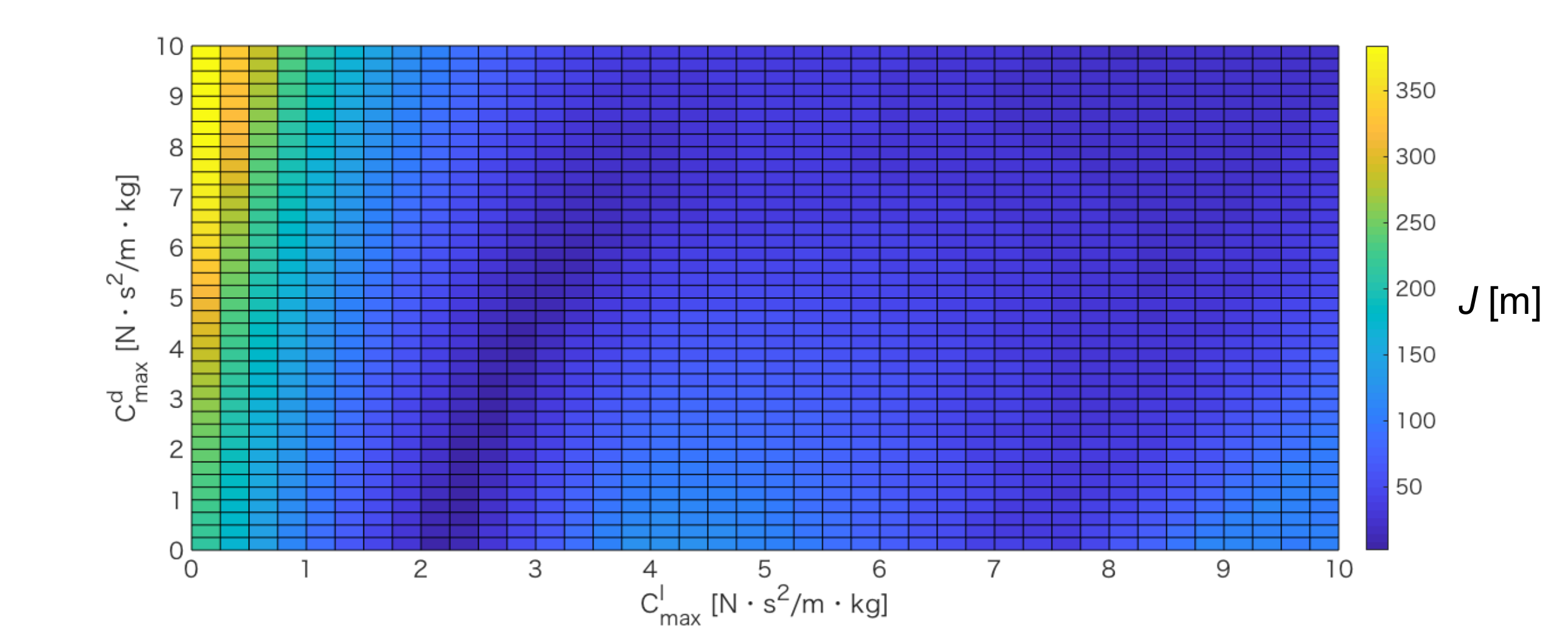}
		\caption{Quantitative analysis on the difference between empirical data and numerical simulation.}
		\label{fig:gridkekka}
	\end{center}
\end{figure}

\begin{figure}[h]
	\begin{center}
		\includegraphics[height=5cm]{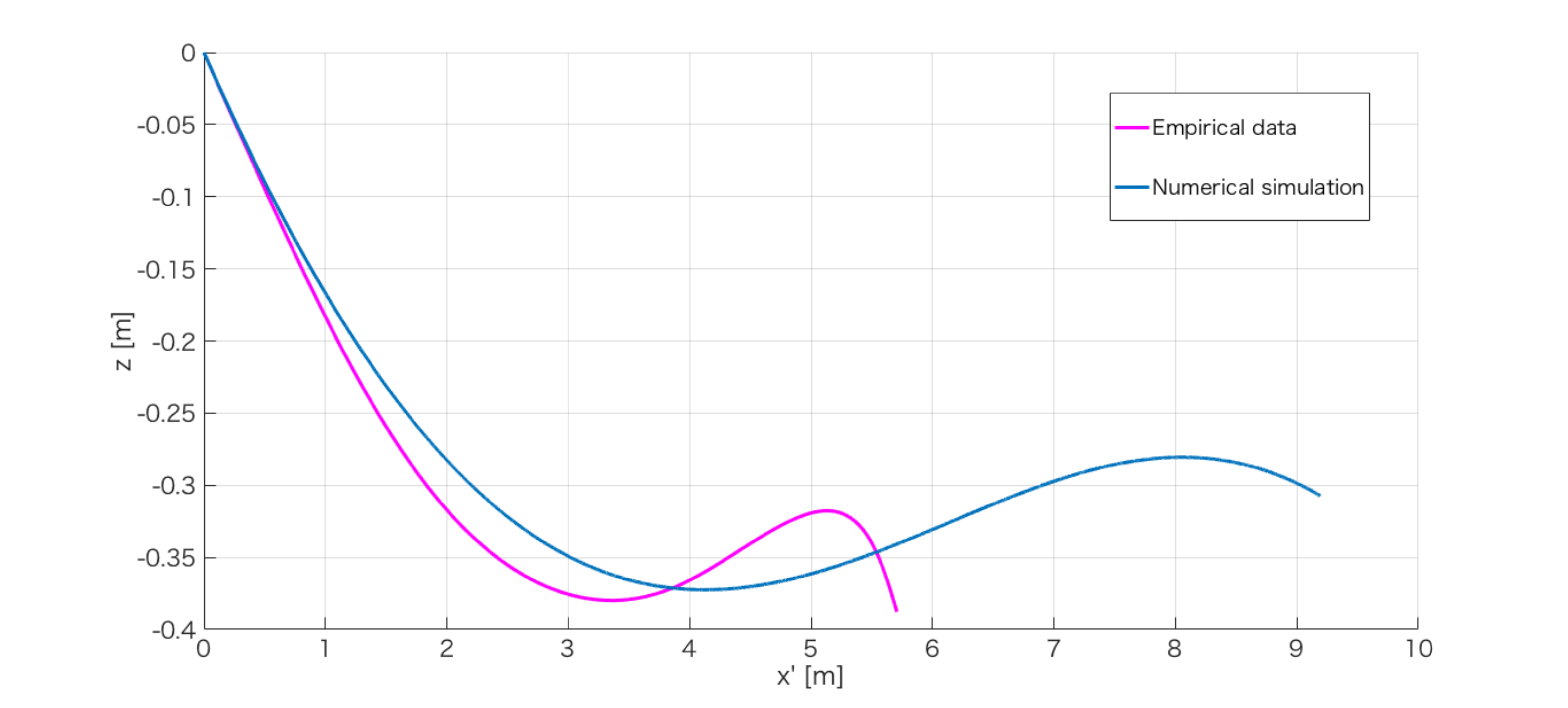}
		\caption{Comparison of flight trajectories between empirical data and numerical simulation.}
		\label{fig:simtra}
	\end{center}
\end{figure}

Figure \ref{fig:gridkekka} shows the distribution of the evaluation function calculated in the range of $0\leq C_{l}^{\rm max}\leq10$ and $0\leq C_{d}^{\rm max}\leq10$.
Here, a larger value of the evaluation function is depicted by yellow, and a smaller value of the evaluation function is depicted by blue.  
Consequently, we found that the set of $(C_{l}^{\rm max}$, $C_{d}^{\rm max})=(2.25, 1.25)$ minimizes the evaluation function.
Figure \ref{fig:simtra} shows the trajectory obtained from the numerical simulation with  $(C_{l}^{\rm max}$, $C_{d}^{\rm max})=(2.25, 1.25)$ (see  a blue line in this figure) that is similar with the trajectory of the empirical data (see a pink line in this figure).

\section{Conclusion and discussion}

In this study, we examined the landing behavior of a seagull based on field observation and mathematical modeling.
In a field observation, we quantified the flight trajectory of an actual seagull and its angle of attack in a three dimensional space by analyzing the video data of two cameras.
Consequently, it is demonstrated that the seagull accomplished a smooth landing on a small target (buoy) while dynamically controlling its angle of attack.
Next, we analyzed a mathematical model describing the flight dynamics of the seagull.
This model assumes that the seagull changes its angle of attack, varies the forces of lift and drag, and controls its flight velocity.
By assuming the empirical data on the angle of attack, we performed numerical simulation of the mathematical model.
It is demonstrated that the flight trajectory of the numerical simulation is consistent with the trajectory of the empirical data, indicating that the control of the angle of attack is essential for an actual seagull to smoothly land on a target while repeating rising up and falling down.

Comparison between numerical simulation and empirical data shows that a seagull exceeded an actual landing point in the case of numerical simulation (see Fig. \ref{fig:simtra}).
This means that the current mathematical model is insufficient to describe the detailed mechanism of the landing behavior.
To solve this problem, for example, we need to consider the effect of wind speed that strongly affects the values of lift and drag.
By measuring the wind speed in an actual field and taking its effect into the mathematical model, it is likely that the flight trajectory of the empirical data can be reproduced more precisely by numerical simulation.
Another problem is that we only used a single data set of a field observation. To robustly show the efficacy of the landing behavior of a seagull, we need to record the landing behavior more times under various environments, and compare the empirical data with the numerical simulation of the improved mathematical model.


\end{document}